\documentstyle[12pt]{article}

\def\llangl{\left\langle}
\def\rrangl{\right\rangle}
\def\lbrk{\left[}
\def\rbrk{\right]}
\begin{document}
\begin{center}
\Large
\bf
Fragmentation model and quark/hadron system
\normalsize
\footnote{Invited talk at the '93 Seoul Workshop on
 ``Strangeness in Nuclear Physics'', Dec. 6-7, 1993, Seoul, Korea.}
\vskip 1.0cm
{\bf Suk-Joon Lee}
\footnote{Electronic address: ssjlee@nms.kyunghee.ac.kr} \\
\vskip 0.3cm
{\it Department of Physics, College of Natural Sciences   \\
 Kyung Hee University, Yongin, Kyungkido \ 449-701, Korea}\\
\end{center}
\vskip 1cm
\begin{abstract}
Nuclear multifragmentation process can be viewed as a recombination
of nucleons into clusters of various sizes.
In a combinatorial analysis,
various moments of cluster size distribution appear to be quite simple
in terms of canonical partition functions.
This simple model can also describe a branching phenomena
and a clusterization phenomena.
The possibility of applying this fragmentation model in describing
a hadronization of quark-gluon system and a strangeness production
is discussed.
\end{abstract}

\section{Introduction}

\hspace*{30pt}
A nuclear system at low excitation energy may be described as a Fermi
liquid of nucleons which are strongly coupled each other through
the mean field interaction.
On the other hand, a hot nuclear system at a temperature higher than
the nuclear binding energy per nucleon may be treated as a Fermi gas.
Within the region where the excitation energy is near to the binding
energy lie the possibility of seeing a liquid-gas phase transition
in a nuclear matter.
However, except in infinite nuclear matter,
the number of nucleons in a nuclear system is too small to be
considered as a thermodynamic system.
Only up to few hundred nucleons are involved in a heavy ion collision.
A finite nuclear system with the excitation energy near to the binding
energy breaks up into many smaller pieces
instead of forming an equilibrated system of either liquid or gas.
This excitation energy region is called the multifragmentation region
and the study of nuclear multifragmentation phenomena requires
nonequilibrium considerations and studies of fluctuation and
correlation due to the finiteness of the system.

Based on a combinatorial analysis,
a simple exactly solvable model describing
this multifragmentation process has been developed \cite{mekjian,general}.
This simple model fits various data in nucleon-nucleus and nucleus-nucleus
collisions \cite{frgft,selfsim}. The fits are as good as the fits using
more complicated fragmentation model based on the thermodynamic
statistical model \cite{pronuc}, Monte-Calro calculation
\cite{bondorf}, and percolation model \cite{perco}.
Considering a control parameter as a function of temperature or time,
this model can describe branching process \cite{hwa} and dynamical
behavior of fragmentation or clusterization \cite{general}.

At much higher excitation energy, nuclear system may transform into
a quark-gluon plasma.
This transformation may be viewed as a fragmentation process
of a nucleon into quarks and gluons.
At low energy, quarks are strongly bound inside of a hadron
while the asymptotic freedom for quarks may be applicable at
a temperature higher than the hadronic binding energy.
On the other hand, as the system cools down,
quarks and gluons clusterize forming colorless hadrons.
Both the formation of quark-gluon system and its hadronization
have a similar structure with the nuclear fragmentation process.

This talk summarizes a fragmentation model (Section \ref{sectfrag})
and its branching and evolution aspects (Section \ref{sectevl})
which are discussed in more detail in Ref.\cite{general}.
In Section \ref{sectqrk}, the possibility of applying the simple
fragmentation model in describing a quark-hadron phase transition
is discussed by considering pions as a simple example.

\section{Fragmentation Model} \label{sectfrag}

\hspace*{30pt}
A method for investigating the fragmentation of a nucleus with $A$
nucleons into pieces of smaller size is developed
\cite{mekjian,general,otherx}.
The partitioning of $A$ initial nucleons into groups of varying
sizes gives rise to a distribution of cluster sizes.
Let $n_k$ be the number of clusters of size $k$.
The $n_1$ is then the number of nucleons not in clusters.
A partition
\begin{eqnarray}
 \vec n = \{n_i\} = (n_1, n_2, n_3, \cdots)    \label{partitn}
\end{eqnarray}
defines a particular arrangement of the initial nucleons
into clusters $(n_2, n_3, \cdots)$ and
monomers ($n_1$) regardless of which nucleon belongs to
which cluster.
The number of clusters in a particular partition is
called the multiplicity $M$;
\begin{eqnarray}
 M = \sum_{i=1}^A n_i .   \label{multip}
\end{eqnarray}
A constraint exists which is $A = \sum_{i=1}^A n_i i$,
i.e., the conservation of the total number of nucleons
in the nuclear multifragmentation process.

Now, let us consider a more general case for the later use
in Sections \ref{sectevl} and \ref{sectqrk}.
Suppose we have a system with $N$ different species, each species,
say the species $k$, appearing $n_k$ times in the system.
When each species $k$ has an additive quantity $\alpha_k$,
the total value of this quantity of the whole system is given by
\begin{eqnarray}
 A = \sum_{i=1}^N n_i \alpha_i   \label{aconstr}
\end{eqnarray}
for a particular partition $\vec n$.
Here $A$ may have any value depending on the $\alpha_i$.
However, when we study the nuclear fragmentation phenomena
in terms of the cluster sizes, $\alpha_i = i$ is the number of
nucleons in each cluster and $A$ is an integer representing the
total number of nucleons in the system.
We can also specify each cluster by their energy, i.e.,
$\alpha_i = E_i$, and then $A = E = \sum_i n_i E_i$ becomes
the total energy of the system.
For $\alpha_i = 1$ as in a jet fragmentation process,
$\alpha_i$ and $A$ are the number of jets and $A = M$.
In the hadron-quark phase transition of nucleons, $\alpha_i$ and $A$
may represent the baryon number of quarks and nucleons.

In a nonequilibrium phenomena, a system may have many different
partitions $\vec n$.
For a collision involving $^{235}$U + $^{235}$U,
the number of possible partitions $\vec n$ is of the order $10^{20}$.
Weighting each species $i$ by parameters $x_i$ and $\beta_i$,
we can assign a weight
\begin{eqnarray}
 G_A(\vec n,\vec x) &=& \frac{\Gamma(A+1)}{M!}
          \lbrk\frac{M!}{n_1! n_2! \cdots n_N!}\rbrk
          \prod_{i=1}^N \lbrk\frac{x_i}{\beta_i}\rbrk^{n_i}
   ~=~ \Gamma(A+1) \prod_{i=1}^N \lbrk\frac{x_i^{n_i}}
                        {\beta_i^{n_i} n_i!}\rbrk      \label{weight}
\end{eqnarray}
to each partition $\vec n$.
Here $\vec x = (x_1, x_2, x_3, \cdots)$ and
the $\Gamma(A+1)$ is the gamma function which is $A!$ for an integer $A$.
The $n_i!$ is the Boltzmann factor arising from the coexistence of
$n_i$ identical species $i$ in the partition.
The parameter $\beta_i$ may be related to the partition function
factor of species $i$ originating from its internal structure and $x_i$
to the partition function factor related to the species $i$ as a whole.
In a nuclear fragmentation with $\alpha_i = i$,
the choice of $\beta_i = i$ counts for
the cyclic rearrangements of the nucleons in a cluster of size $i$,
and $x_i \propto e^{-F_i(T,V)/k_BT}$ with the free energy $F_i(T,V)$ of
the cluster $i$ at a temperature $T$ and a freeze out volume $V$.

Summing the weight Eq.(\ref{weight}) over all possible partitions
with a fixed $A$,
the weight for all the partitions having a fixed value $A$ becomes
\begin{eqnarray}
 Q_A(\vec x) &=& \sum_{\{n_i\}_A} G_A(\vec n,\vec x)
   ~=~ \sum_{\{n_i\}_A} \Gamma(A+1)
        \prod_{i=1}^N \lbrk\frac{x_i^{n_i}}{\beta_i^{n_i} n_i!}\rbrk .
            \label{qax}
\end{eqnarray}
This $Q_A(\vec x)$ is then the canonical partition function of
the system with a fixed value $A$.
For the case of an integer $A$ such as in a nuclear fragmentation, we can
get easily the canonical partition function from a generating function
\begin{eqnarray}
 Q(u,\vec x) &=& \sum_{\{n_i\}} G_A(\vec n,\vec x) \frac{u^A}{\Gamma(A+1)}
   ~=~ \sum_{M=0}^\infty \frac{1}{M!} \lbrk \sum_{i=1}^N
          \frac{u^{\alpha_i}}{\beta_i} x_i \rbrk^M
   ~=~ \exp\lbrk\sum_{i=1}^N \frac{1}{\beta_i} x_i u^{\alpha_i} \rbrk
        \label{generqux}
\end{eqnarray}
which is an infinite sum of multinomials.
By summing the weight $G_A(\vec n, \vec x)$ over the
partitions with a fixed $A$ and a fixed $n_k$ of species $k$,
we obtain the weight
\begin{eqnarray}
 Q_A^{n_k}(\vec x) &=& \sum_{\{n_i\}_{A,n_k}} G_A(\vec n, \vec x)
   ~=~ \sum_{\{n_i\}_{A,n_k}} \Gamma(A+1)
        \prod_{i=1}^N \lbrk\frac{x_i^{n_i}}{\beta_i^{n_i} n_i!}\rbrk
           \label{qank}
\end{eqnarray}
for all the partitions having
$n_k$ species of type $k$ with a fixed value of $A$.

The probability for a specific partition $\vec n$ to appear in the
canonical ensemble ($Q_A$) of a fixed value $A$ is given by
\begin{eqnarray}
 P_A(\vec n , \vec x) &=& \frac{G_A(\vec n , \vec x)}{Q_A(\vec x)}
   ~=~ \left( \Gamma(A+1) \prod_{i=1}^N \lbrk\frac{x_i^{n_i}}
             {\beta_i^{n_i} n_i!}\rbrk \right) \left/
   \left( \sum_{\{n_i\}_A} \Gamma(A+1) \prod_{i=1}^N \lbrk\frac{x_i^{n_i}}
          {\beta_i^{n_i} n_i!}\rbrk \right) \right. .    \label{pan}
\end{eqnarray}
We also define the inclusive multiplicity distribution
\begin{eqnarray}
 P_A(n_k , \vec x) &=& \sum_{\{n_i\}_{A,n_k}} P_A(\vec n , \vec x)
    ~=~ \frac{Q_A^{n_k}(\vec x)}{Q_A(\vec x)}      \label{pank}
\end{eqnarray}
as the probability of having $n_k$ species of type $k$ in the
canonical ensemble.

Due to the $n_i$ dependence of the weight $G_A(\vec n, \vec x)$
with the specific form of $n_i!$ and $(x_i/\beta_i)^{n_i}$,
the derivatives of Eq.(\ref{weight}) lead back to the weight
$G_A(\vec n, \vec x)$ itself with a constant factor;
\begin{eqnarray}
 \lbrk x_j \frac{d}{d x_j}\rbrk^l \lbrk x_k \frac{d}{d x_k}\rbrk^m
    G_A(\vec n , \vec x)
    ~=~ n_j^l n_k^m G_A(\vec n , \vec x) ,   \hspace{5.8cm}
       \label{xdxdg}  \\
 \lbrk x_j \rbrk^l \lbrk x_k\rbrk^m \lbrk \frac{d}{dx_j}\rbrk^l
    \lbrk \frac{d}{d x_k} \rbrk^m G_A(\vec n , \vec x)
   ~=~ \frac{n_j!}{(n_j-l)!} \frac{n_k!}{(n_k-m)!} G_A(\vec n , \vec x)
           \hspace{2.5cm}          \nonumber \\
   ~=~ \frac{\Gamma(A+1)}{\Gamma(A+1 - \{m\alpha_k + l\alpha_j\})}
       \lbrk\frac{x_k}{\beta_k}\rbrk^m \lbrk\frac{x_j}{\beta_j}\rbrk^l
       G_{A - \{m\alpha_k + l\alpha_j\}}({\vec n}', \vec x) ,
           \label{dxdxdg}
\end{eqnarray}
where ${\vec n}'$ is used to discriminate partitions having
$\sum_i n'_i \alpha_i = (A - \{m\alpha_k + l\alpha_j\})$
from the partitions $\vec n$ with $A$.
Notice that Eq.(\ref{xdxdg}) becomes zero only when
$n_j$ or $n_k$ is zero in contrast to Eq.(\ref{dxdxdg})
which is non-zero only when $l \le n_j$ and $ m \le n_k$.
Eq.(\ref{dxdxdg}) also applies to the case of $j = k$
with $n_j = (n_k - m)$, i.e., replacing the factorial factors
on the right hand side by $n_k!/(n_k - m - l)!$.
Due to the form of Eqs.(\ref{xdxdg}) and (\ref{dxdxdg}),
it is quite simple to consider various average values in the canonical
ensemble ($Q_A$) of a fixed value $A$ with the corresponding probability
of Eq.(\ref{pan}):
\begin{eqnarray}
 \llangl \frac{n_k!}{(n_k-m)!} \frac{n_j!}{(n_j-l)!} \cdots \rrangl
  ~=~ \sum_{\{n_i\}_A} P_A(\vec n, \vec x) \lbrk\frac{n_k!}{(n_k-m)!}\rbrk
        \lbrk\frac{n_j!}{(n_j-l)!}\rbrk \cdots
            \hspace{1.0in}     \nonumber  \\
  ~=~ \left\{ \lbrk\frac{x_k}{\beta_k}\rbrk^m
       \lbrk\frac{x_j}{\beta_j}\rbrk^l \cdots \right\}
       \lbrk\frac{Q_{A - \{m \alpha_k + l \alpha_j + \cdots\}}(\vec x)}
            {\Gamma(A+1 - \{m \alpha_k + l \alpha_j + \cdots\})}\rbrk
  \left/ \lbrk\frac{Q_A(\vec x)}{\Gamma(A+1)}\rbrk \right.  \nonumber  \\
  ~=~ Y_A^m(k,\vec x) ~Y_{A-m \alpha_k}^l(j,\vec x) \cdots
   ~~=~~ Y_{A-l \alpha_j}^m(k,\vec x) ~ Y_A^l(j,\vec x) \cdots .
            \hspace{0.82in}
                \label{nkmnjl}
\end{eqnarray}
Here $\langle ~\rangle$ represents an average over the canonical
ensemble $Q_A(\vec x)$ and
\begin{eqnarray}
 Y_A^m(k,\vec x)
    &=& \sum_{n_k=0}^\infty \frac{n_k!}{(n_k-m)!} P_A(n_k,\vec x)
    ~=~ \lbrk\frac{x_k}{\beta_k}\rbrk^m
        \frac{\Gamma(A+1)}{\Gamma(A+1 - m \alpha_k)}
        \frac{Q_{(A - m \alpha_k)}(\vec x)}{Q_A(\vec x)}
                 \label{yamk}
\end{eqnarray}
is the $m$-th order moment of $n_k$.
Eq.(\ref{nkmnjl}) also applies for $j = k$ when $n_j = n_k - m$.
For this case, Eq.(\ref{nkmnjl}) becomes Eq.(\ref{yamk}) with the $m$
of Eq.(\ref{yamk}) set equal to $m + l$.
The inclusive multiplicity distribution $P_A(n_k, \vec x)$
of a species $k$ in a canonical ensemble can be looked at through a
quantity defined by
\begin{eqnarray}
 F_{A,m}^{n_k}(\vec x)
  &=& \frac{n_k!}{(n_k - m)!} Q_A^{n_k}(\vec x)
  ~= \sum_{\{n_i\}_{A,n_k}}\! \frac{n_k!}{(n_k-m)!} G_A(\vec n, \vec x)
  ~=~ \frac{\Gamma(A+1)}{\Gamma(A+1 - m \alpha_k)}
      \tilde F_{A,m}^{n_k} (\vec x)  \hspace{1.0cm}       \label{fam}
\end{eqnarray}
for $m \le n_k$, which is related with the moments through
 $Y_A^m(k, \vec x) = \left. \sum_{n_k = 0}^\infty F_{A,m}^{n_k}(\vec x)
     \right/ Q_A(\vec x)$.

Once we know the canonical partition function $Q_A(\vec x)$
of a canonical ensemble with a fixed $A$,
we can easily obtain various moments of the species distribution $n_k$
by Eq.(\ref{nkmnjl}).
Specifically, the choice of $\alpha_i = \beta_i = i$ leads to various
simple exactly solvable canonical systems \cite{general,otherx}.
When all the species have a same weight $x_i = x$, the canonical
partition function simply becomes
 $Q_A(\vec x) = \Gamma(x + A) / \Gamma(x)$.
This simple exactly solvable model describes the average behavior of
nuclear multifragmentation phenomena very well
\cite{general,frgft,selfsim,otherx,canxy}.
The model were also able to describe fluctuations, correlations,
intermittency, power law and scaling behavior involved in the nuclear
multifragmentation phenomena \cite{general,selfsim,geomet,emuls}.
There are various sets of $\vec x$ each of which gives an exactly
solvable model \cite{otherx}. A specific example among them is the case
of $x_i = (1 - q^i)^{-1}$ which can describe the even-odd oscillating
behavior appearing in various fields such as a pairing effect in a nucleus
or the size of parties reserving a table in a restaurant \cite{otherx}.

\section{Evolution and Branching Phenomena} \label{sectevl}

\hspace*{30pt}
A particular set of $x_i$ values gives a specific distribution of
possible partitions through Eq.(\ref{pan}) which in turn determines
uniquely the various moments of the multiplicity $n_k$.
Now suppose the weight parameter $x_i$'s are functions of an evolution
parameter $t$ such as time or temperature, i.e., $x_i = x_i(t)$.
A change of the evolution parameter $t$ of the system induces
a change in the $x_i(t)$'s which assign a different weight given
by $G_A(\vec n,\vec x)$ of Eq.(\ref{weight}) for different partition
$\vec n$. This change in the weight occurring in response to the change
of the evolution parameter $t$ of the system gives the $t$-dependence of
the partition function $Q_A(\vec x)$ of Eq.(\ref{qax})
which describes the evolution of the system in $t$, i.e., how
the average distribution $\llangl n_k\rrangl$  of species $k$ in
a canonical ensemble having a fixed value of $A$ changes with $t$.

The evolution of the canonical system can also be viewed as
a redistribution of the quantity $A = \sum_i n_i \alpha_i$
into various species $k$.
The distribution of $m_k = n_k \alpha_k$ can be represented by
\begin{eqnarray}
 N_k^m(A) &=& \alpha_k^m Y_A^m(k,\vec x)
    ~=~ \alpha_k^m \left. \sum_{n_k = 0}^\infty F_{A,m}^{n_k}(\vec x)
     \right/ Q_A(\vec x) .   \label{nkm}
\end{eqnarray}
(In Ref.\cite{general}, $N_k^m(A)$ was defined by
$\alpha_k Y_A^m(k, \vec x)$.)
For the case of $\alpha_i = i$,
$m_k = n_k k$ is the mass distribution in a partition and
$N_k^1(A) = \llangl m_k \rrangl = k\llangl n_k \rrangl$ can
be considered as the total average number of nucleons belonging to
clusters of size $k$ in the fragmentation of a nucleus of size $A$.
The quantity $N_k^1(A)/A = (k/A) \llangl n_k\rrangl$
is considered in Ref.\cite{otherx} as a probability.
In a nuclear multifragmentation process described in terms of
the cluster sizes ($\alpha_i = i$),
the evolution of a canonical system corresponds
to the redistribution of nucleons into various size of clusters.
On the other hand, when $\alpha_i = E_i$ is the energy of a cluster $i$,
the evolution is the redistribution of energy of the system
into various clusters.
Using Eq.(\ref{dxdxdg}) and
 $\lbrk\frac{d}{d t}\rbrk = \sum_{k=1}^N \lbrk\frac{d}{d t}
      \ln x_k(t) \rbrk \lbrk x_k \frac{d}{d x_k} \rbrk$,
and separating $j < k$ and $j > k$,
we obtain the dynamic equation for $N_k^m(A)$ as
\begin{eqnarray}
 \frac{d}{d t} N_k^m(A)
  &=& \sum_{l=1}^{k-1} J_{(k-l)}^l (A,k,m)
     + \sum_{l=1}^{N-k} J_{(k+l)}^{-l} (A,k,m) ,
          \label{dtdnam}
\end{eqnarray}
where
\begin{eqnarray}
 J_j^l(A,k,m)
     &=& C_j^l \lbrk
           \llangl\frac{\alpha_k^m n_k!}{(n_k - m)!} \alpha_j n_j\rrangl
         - \llangl\frac{\alpha_k^m n_k!}{(n_k - m)!}\rrangl
              \llangl\alpha_j n_j\rrangl \rbrk   \nonumber  \\
     &=& C_j^l \lbrk N_k^m(A - \alpha_j) N_j^1(A)
                   - N_k^m(A) N_j^1(A) \rbrk ,    \label{jjl}  \\
 C_k^l ~=~ - C_{k+l}^{-l}
       &=& \lbrk\frac{1}{\alpha_k} \left(\frac{d}{dt}\right) \ln x_k(t)
         - \frac{1}{\alpha_{(k+l)}} \left(\frac{d}{dt}\right)
             \ln x_{(k+l)}(t) \rbrk .      \label{ckl}
\end{eqnarray}
Eqs.(\ref{dtdnam}) and (\ref{jjl}) show
that the changing rate of $N_k^m(A)$
depends not on the mean behavior but on fluctuations
and correlations in the distribution of $m_k = n_k \alpha_k$.
The distribution is in equilibrium when $J_j^l(A,k,m) = 0$, i.e.
when $C_k^l = 0$ or when there are no fluctuations nor correlations.

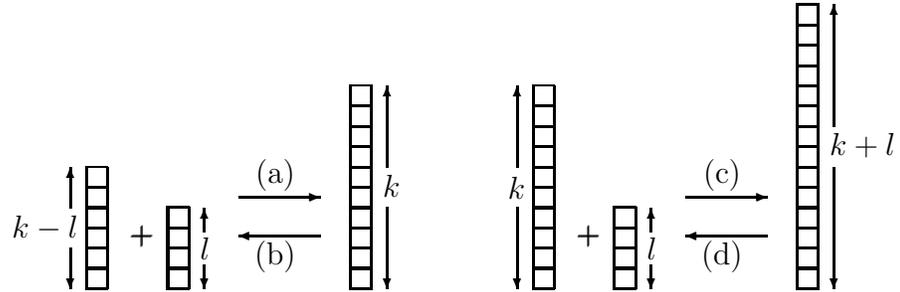
\begin{figure}[phtb]
\begin{center}

\setlength{\unitlength}{0.54cm}
\begin{picture}(21.5,7.2)(1.0,0.3)

\newsavebox{\diagbox}
\savebox{\diagbox}(0.5,0.5){ }

\multiput(2.5, 1.0)(0,0.5){6}{\frame{\usebox{\diagbox}}}
\multiput(4.5, 1.0)(0,0.5){4}{\frame{\usebox{\diagbox}}}
\multiput(9.0, 1.0)(0,0.5){10}{\frame{\usebox{\diagbox}}}
\put(2.5, 4.01){\line(1,0){0.5}}
\put(2.5, 1.01){\line(1,0){0.5}}
\put(2.51, 1.0){\line(0,1){3.0}}
\put(3.03, 1.0){\line(0,1){3.0}}
\put(4.5, 3.01){\line(1,0){0.5}}
\put(4.5, 1.01){\line(1,0){0.5}}
\put(4.51, 1.0){\line(0,1){2.0}}
\put(5.03, 1.0){\line(0,1){2.0}}
\put(9.0, 6.01){\line(1,0){0.5}}
\put(9.0, 1.01){\line(1,0){0.5}}
\put(9.01, 1.0){\line(0,1){5.0}}
\put(9.52, 1.0){\line(0,1){5.0}}
\multiput(2.1,2.0)(0.01,0){2}{\vector(0,-1){1.0}}
\multiput(2.1,3.0)(0.01,0){2}{\vector(0,1){1.0}}
\put(2.0,2.2){\makebox(0.3,0.6)[r]{$k - l$}}
\multiput(5.4,1.6)(0.01,0){2}{\vector(0,-1){0.6}}
\multiput(5.4,2.4)(0.01,0){2}{\vector(0,1){0.6}}
\put(5.3,1.7){\makebox(0.3,0.6)[l]{$l$}}
\multiput(9.9,3.0)(0.01,0){2}{\vector(0,-1){2.0}}
\multiput(9.9,4.0)(0.01,0){2}{\vector(0,1){2.0}}
\put(9.8,3.25){\makebox(0.3,0.6)[l]{$k$}}
\multiput(3.7,2.0)(0.02,0.02){2}{\makebox(0.3,0.6)[c]{$+$}}
\multiput(6.25,3.25)(0,0.01){2}{\vector(1,0){2.0}}
\put(7.0,3.5){\makebox(0.3,0.6)[c]{(a)}}
\multiput(8.25,2.3)(0,0.01){2}{\vector(-1,0){2.0}}
\put(7.0,1.5){\makebox(0.3,0.6)[c]{(b)}}
\multiput(13.5, 1.0)(0,0.5){10}{\frame{\usebox{\diagbox}}}
\multiput(15.5, 1.0)(0,0.5){4}{\frame{\usebox{\diagbox}}}
\multiput(20.0, 1.0)(0,0.5){14}{\frame{\usebox{\diagbox}}}
\put(13.5, 6.01){\line(1,0){0.5}}
\put(13.5, 1.01){\line(1,0){0.5}}
\put(13.51, 1.0){\line(0,1){5.0}}
\put(14.03, 1.0){\line(0,1){5.0}}
\put(15.5, 3.01){\line(1,0){0.5}}
\put(15.5, 1.01){\line(1,0){0.5}}
\put(15.51, 1.0){\line(0,1){2.0}}
\put(16.03, 1.0){\line(0,1){2.0}}
\put(20.0, 8.01){\line(1,0){0.5}}
\put(20.0, 1.01){\line(1,0){0.5}}
\put(20.01, 1.0){\line(0,1){7.0}}
\put(20.53, 1.0){\line(0,1){7.0}}
\multiput(13.1,3.0)(0.01,0){2}{\vector(0,-1){2.0}}
\multiput(13.1,4.0)(0.01,0){2}{\vector(0,1){2.0}}
\put(13.0,3.2){\makebox(0.3,0.6)[r]{$k$}}
\multiput(16.4,1.6)(0.01,0){2}{\vector(0,-1){0.6}}
\multiput(16.4,2.4)(0.01,0){2}{\vector(0,1){0.6}}
\put(16.3,1.7){\makebox(0.3,0.6)[l]{$l$}}
\multiput(20.9,4.0)(0.01,0){2}{\vector(0,-1){3.0}}
\multiput(20.9,5.0)(0.01,0){2}{\vector(0,1){3.0}}
\put(20.8,4.25){\makebox(0.3,0.6)[l]{$k + l$}}
\multiput(14.7,2.0)(0.02,0.02){2}{\makebox(0.3,0.6)[c]{$+$}}
\multiput(17.25,3.25)(0,0.01){2}{\vector(1,0){2.0}}
\put(18.0,3.5){\makebox(0.3,0.6)[c]{(c)}}
\multiput(19.25,2.3)(0,0.01){2}{\vector(-1,0){2.0}}
\put(18.0,1.5){\makebox(0.3,0.6)[c]{(d)}}

\end{picture}

\caption[ ]{
Evolution of fragmentation process in the fragmentation
block diagram representation \cite{otherx}.
Each box represents a nucleon and the stack of boxes represents
a cluster.
 } \label{figbox}

\end{center}
\end{figure}

While Eq.(\ref{dtdnam}) seems complex, this equation has a
simple interpretation.
The first term of Eq.(\ref{dtdnam}) represents the creation
of species $k$ arising from the fusion of one of the $N_{(k-l)}^1(A)$
species $(k-l)$ and one of species $l$
when the $J_{(k-l)}^l$ is positive
(process (a) in Fig.\ref{figbox} which is shown for a nuclear
fragmentation process).
If the $J_{(k-l)}^l$ is negative, we can also interpret this same
term as an annihilation process (b) of one of $N_k^1(A)$ species $k$.
The second term of Eq.(\ref{dtdnam}) represents the formation
of species $(k+l)$ arising from the fusion of species $k$ and $l$
(process (c) in Fig.\ref{figbox}), which thus reduces species $k$,
when $J_k^l$ is negative.
When $J_k^l$ is positive, this second term represents the formation
of species $k$ arising from the fission of species $(k+l)$
(process (d) in Fig.\ref{figbox}).

For the $m = 1$ case, Eq.(\ref{dtdnam}) represents the
evolution of the average distribution
 $\llangl m_k\rrangl = \alpha_k \llangl n_k\rrangl$
of species $k$ \cite{general};
\begin{eqnarray}
 \frac{d}{d t} N_k^1(A)
   &=& \sum_{l=1}^{k-1} D_{(k-l)}^l N_{(k-l)}^1(A)
    - \lbrk \sum_{l=1}^{N-k} D_k^l + \sum_{l=1}^{k-1} E_k^l
             \rbrk N_k^1(A)
      + \sum_{l=1}^{N-k} E_{(k+l)}^l N_{(k+l)}^1(A) ,
          \label{dtdnk1}
\end{eqnarray}
where
\begin{eqnarray}
 D_j^l ~=~ C_j^l N_{(j+l)}^1(A - \alpha_j) ,  \hspace{2cm}
 E_j^l ~=~ C_{(j-l)}^l N_{(j-l)}^1(A) .       \label{djlejl}
\end{eqnarray}
The result of Eq.(\ref{dtdnk1}) can be related to
the Chapman-Kolmogorov equation \cite{hwa}
in a continuum limit of the species index $j$ \cite{general}.
Considering only $l = 1$ in Eq.(\ref{dtdnk1}), the result is
further reduced to a Fokker-Planck equation \cite{hwa,fokkplk}.
Due to the discrete value of $k$, Eq.(\ref{dtdnk1}) is applicable
for a finite system with the finite number of $k$ values.

The evolution of the inclusive multiplicity distribution
$P_A(n_k, \vec x)$ of a species $k$ in a canonical ensemble can be looked
at through a change of $F_{A,m}^{n_k}(\vec x)$ defined by Eq.(\ref{fam}).
The evolution of the $F_{A,m}^{n_k}(\vec x)$ contains
information regarding the underlying dynamics such as
in various cascade phenomena.
Using Eq.(\ref{dxdxdg}), we obtain the dynamic equation
for the function $F_{A,m}^{n_k}(\vec x)$ of Eq.(\ref{fam});
\begin{eqnarray}
 \lbrk\frac{d}{d t}\rbrk \! F_{A,m}^{n_k} (\vec x)
   &=& \frac{A}{\alpha_k} \! \lbrk\frac{d}{d t} \ln x_k(t)\rbrk
        \! F_{A,m}^{n_k} (\vec x)
    + \frac{1}{\alpha_k} \sum_{j=1}^N \! \lbrk\frac{d}{d t}
          \ln \frac{x_j^{\alpha_k}(t)}{x_k^{\alpha_j}(t)} \! \rbrk
       \! \lbrk\frac{x_j}{\beta_j}\rbrk \! \frac{A!}{(A - \alpha_j)!}
        F_{(A - \alpha_j),m}^{n_k}(\vec x) .  \hspace{0.8cm}  \label{dfdt}
\end{eqnarray}
This result may be considered as a dynamical equation for
the factorial distribution of species of type $k$
in a canonical ensemble.
The first term represents the dependence on the multiplicity $n_k$ of
the same species $k$
and the second term represents the dependence on the multiplicity
$n_j$ of species of type $j \ne k$.
The Eq.(\ref{dfdt}) for $m = 1$ is the changing rate of the
number of species of type $k$,
$n_k = \left. F_{A,1}^{n_k}(\vec x) \right/ Q_A^{n_k}(\vec x)$,
when each species $i$ has the weight $x_i(t)$.

\begin{figure}[phtb]
\begin{center}

\setlength{\unitlength}{0.5cm}
\begin{picture}(21,9)(0.0,0.0)
\thicklines

\put(2.0,4.45){\makebox(0.3,0.1)[r]{{\bf Initial :} \  $n_k$--jets \
        $\left\{ \makebox(0.2,4.0)[c]{ } \right.$}}
\put(2.5, 8.00){\line(1,0){2.0}}
\put(2.5, 5.50){\line(1,0){2.0}}
\put(3.5, 3.40){\makebox(0.3,0.6)[c]{$\vdots$}}
\put(3.5, 2.60){\makebox(0.3,0.6)[c]{$\vdots$}}
\put(2.5, 1.00){\line(1,0){4.0}}
\put(4.5, 8.00){\line(5,1){4.0}}
\put(4.5, 8.00){\line(5,-1){2.0}}
\put(4.5, 8.00){\circle*{0.3}}
\put(6.5, 7.60){\circle*{0.3}}
\put(6.5, 7.60){\line(5,1){2.0}}
\put(6.5, 7.60){\line(5,-1){2.0}}
\put(4.5, 5.50){\line(5,1){4.0}}
\put(4.5, 5.50){\line(5,-1){4.0}}
\put(4.5, 5.50){\circle*{0.3}}
\put(6.5, 1.00){\line(5,1){2.0}}
\put(6.5, 1.00){\line(5,-1){2.0}}
\put(6.5, 1.00){\circle*{0.3}}
\put(9.7, 4.00){\makebox(0.3,0.6)[l]{$\cdots\cdots$}}
\put(12.5, 7.60){\line(5,1){2.0}}
\put(14.5, 8.00){\circle*{0.3}}
\put(14.5, 8.00){\line(5,1){4.0}}
\put(14.5, 8.00){\line(5,-1){2.0}}
\put(16.5, 7.60){\circle*{0.3}}
\put(16.5, 7.60){\line(5,1){2.0}}
\put(16.5, 7.60){\line(5,-1){2.0}}
\put(12.5, 6.00){\line(1,0){4.0}}
\put(16.5, 6.00){\circle*{0.3}}
\put(16.5, 6.00){\line(5,1){2.0}}
\put(16.5, 6.00){\line(5,-1){2.0}}
\put(17.5, 4.20){\makebox(0.3,0.6)[r]{$\vdots$}}
\put(12.5, 3.00){\line(1,0){6.0}}
\put(12.5, 1.80){\line(5,-1){2.0}}
\put(14.5, 1.40){\circle*{0.3}}
\put(14.5, 1.40){\line(5,1){2.0}}
\put(16.5, 1.80){\circle*{0.3}}
\put(16.5, 1.80){\line(5,1){2.0}}
\put(16.5, 1.80){\line(5,-1){2.0}}
\put(14.5, 1.40){\line(5,-1){4.0}}
\put(19.0,4.65){\makebox(0.3,0.1)[l]{$\left. \makebox(0.1,4.8)[c]{ }
               \right\}$  \  $A$--jets \ {\bf : Final}}}

\end{picture}

\caption[ ]{
Branching diagram \cite{hwa}.
Each vertex (solid circle) represents a binary branching of
Eq.(\ref{furry}).
The $n_k$ initial jets branch into the final $A$ jets at
($A - n_k$)--vertices.
 } \label{figbrnch}

\end{center}
\end{figure}
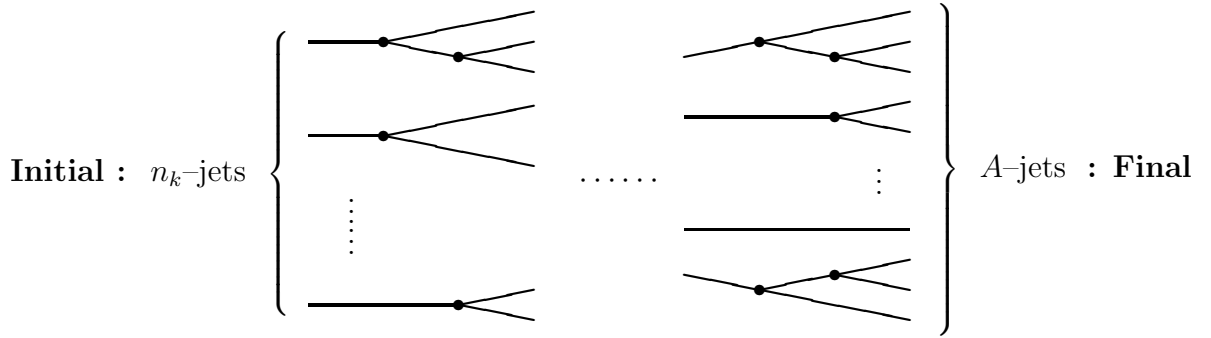
An interesting application of Eq.(\ref{dfdt}) is to
branching processes and jet fragmentation.
For $\alpha_i = \beta_i = 1$, in which
$Q_A(\vec x) = (\sum_{i=1}^N x_i)^A
 = \sum_{n_k} \tilde F_{A,0}^{n_k}(\vec x)$
is a simple multinomial \cite{general} with $A = M$,
Eq.(\ref{dfdt}) reduces to
\begin{eqnarray}
 \frac{d}{d t} \tilde F_{A,m}^{n_k} (\vec x)
   &=& A \lbrk\frac{d}{d t} \ln x_k\rbrk \tilde F_{A,m}^{n_k} (\vec x)
     + (A-m) \sum_{j=1}^N \lbrk\frac{d}{d t}
         \ln \left(\frac{x_j}{x_k}\right)\rbrk
         x_j \tilde F_{(A-1),m}^{n_k} (\vec x) .    \label{dfdtab1}
\end{eqnarray}
Furthermore, if $N = 2$ with $x_1(t) = 1/\omega(t)$ and
$x_2(t) = 1 - 1/\omega(t)$, then
the Eq.(\ref{dfdtab1}) is reduced to a generalization of a dynamical
equation ($m = 1$) for the multiplicity distribution function
of jet fragmentation in $\phi^3$ theory \cite{hwa}
having three lines at each vertex.
For the species $k = 1$ with $m = 1$,
the solution of Eq.(\ref{dfdtab1}) becomes the Furry distribution
\begin{eqnarray}
 \tilde F_{A,1}^{n_k}(\omega)
    &=& \frac{\Gamma(A)}{\Gamma(n_k)\Gamma(A-n_k+1)}
        \left(\frac{1}{\omega}\right)^{n_k}
        \left(1 - \frac{1}{\omega}\right)^{A-n_k}  \label{furry}
\end{eqnarray}
in $\phi^3$ theory
with the evolution parameter $\omega(t) = e^{\beta t}$.
Here the $A$ is the number of final jet lines at $t$
and $n_k$ is the number of initial jet lines at $t = 0$.
In $\phi^3$ theory, each initial line (stem) branches into
$N = 2$ new lines (branches) at each step (Fig.\ref{figbrnch}).
Here the species $k = 1$ is the old jet and the species $j = 2$
represents the new born jet from the old one with the probability
of $x_2 = 1 - 1/\omega$.
There are $(A - n_k)$ branching vertices.
For a given $A$ final jets, the $n_k$ initial jets should branch
$(A - n_k)$ times producing $(A - n_k)$ new jets.
$\tilde F_{A,1}^{n_k}$ is the probability of starting from $n_k$
initial jets to have final jets of a fixed number $A$.
Detail discussions concerning the relation between the Furry
distribution and branching processes can be found in Ref.\cite{hwa}.

\section{Quark-Hadron System}  \label{sectqrk}

\hspace*{30pt}
Study of quark-hadron phase transition is more complicated than
the study of nuclear multifragmentation process.
In a nuclear multifragmentation region, the nuclear system can be
considered as a system constituted with one type of elements (nucleons)
and the total number $A$ of nucleons is conserved.
In contrast to this, hadrons are constituted with quarks having
different colors and flavors.
The constituents carry color but the hadrons are colorless.
We also need to consider anti-particles in this energy region
thus having a non-conserved total number of particles.
Further investigations are needed in order to use the fragmentation
model to describe a hadron-quark phase transition.
Here we just discuss how the fragmentation model may
look like when we apply it to pions (simplest hadron).

\begin{figure}[bhtp]
\begin{center}

\def\fynlnth{0.5cm}
\def\fynlnths{0.4cm}
\setlength{\unitlength}{\fynlnth}
\begin{picture}(28.0,8)(0.0,0.0)
\thicklines

\newcommand{\ovalh}{\begin{picture}(0.9,0.6)(0.0,0.0) \thicklines
   \put(0.40,0.0){\oval(0.8,0.8)[b]}
   \put(0.65,0.0){\oval(0.3,0.3)[t]}
        \end{picture}}
\newcommand{\ovalu}{\begin{picture}(0.9,0.6)(0.0,0.0) \thicklines
   \put(0.40,0.0){\oval(0.8,0.8)[b]}
   \put(0.80,0.0){\line(0,1){0.25}}
   \put(0.65,0.25){\oval(0.3,0.3)[t]}
        \end{picture}}
\newcommand{\ovald}{\begin{picture}(0.9,0.6)(0.0,0.0) \thicklines
   \put(0.40,0.0){\oval(0.8,0.8)[b]}
   \put(0.65,0.0){\oval(0.3,0.3)[tr]}
   \put(0.65,-0.25){\oval(0.3,0.8)[tl]}
        \end{picture}}
\newcommand{\pionh}{\begin{picture}(0.9,0.6)(0.0,0.0) \thicklines
   \multiput(0.0,0.0)(0.4,0.0){7}{\circle*{0.15}}
        \end{picture}}
\newcommand{\pionu}{\begin{picture}(0.9,0.6)(0.0,0.0) \thicklines
   \multiput(0.0,0.0)(0.4,0.2){8}{\circle*{0.15}}
        \end{picture}}
\newcommand{\piond}{\begin{picture}(0.9,0.6)(0.0,0.0) \thicklines
   \multiput(0.0,0.0)(0.4,-0.2){8}{\circle*{0.15}}
        \end{picture}}
\newcommand{\glonh}{\setlength{\unitlength}{\fynlnth}
        \begin{picture}(0.9,0.6)(0.0,0.0) \thicklines
   \put(0.00,0.0){\line(1,0){1.0}}
   \put(0.85,0.0){\oval(0.3,0.6)[tr]}
   \put(0.85,0.15){\oval(0.3,0.3)[tl]}
   \multiput(0.7,0.15)(0.5,0){3}{\ovalh}
   \put(2.6,0.15){\oval(0.8,0.8)[b]}
   \put(2.85,0.15){\oval(0.3,0.3)[tr]}
   \put(2.85,0.0){\oval(0.3,0.6)[tl]}
   \put(2.7,0.0){\line(1,0){1.0}}
        \end{picture}}
\newcommand{\glonu}{\setlength{\unitlength}{\fynlnths}
        \begin{picture}(0.9,0.6)(0.0,0.0) \thicklines
   \put(0.00,0.0){\line(2,1){1.0}}
   \put(0.85,0.5){\oval(0.3,0.5)[tr]}
   \put(0.85,0.6){\oval(0.3,0.3)[tl]}
   \multiput(0.7,0.6)(0.5,0.25){3}{\ovalu}
   \put(2.6,1.35){\oval(0.8,0.8)[b]}
   \put(2.85,1.35){\oval(0.3,0.8)[tr]}
   \put(2.85,1.25){\oval(0.3,1.0)[tl]}
   \put(2.7,1.25){\line(2,1){1.0}}
        \end{picture}}
\newcommand{\glond}{\setlength{\unitlength}{\fynlnths}
        \begin{picture}(0.9,0.6)(0.0,0.0) \thicklines
   \put(0.0,0.0){\line(2,-1){1.0}}
   \put(0.85,-0.5){\oval(0.3,1.0)[tr]}
   \put(0.85,-0.4){\oval(0.3,0.8)[tl]}
   \multiput(0.7,-0.4)(0.5,-0.25){3}{\ovald}
   \put(2.6,-1.15){\oval(0.8,0.8)[b]}
   \put(2.85,-1.15){\oval(0.3,0.3)[tr]}
   \put(2.85,-1.25){\oval(0.3,0.5)[tl]}
   \put(2.7,-1.25){\line(2,-1){1.0}}
        \end{picture}}
\put(1.0,7.0){\pionh}
\put(0.2,6.75){\makebox(0.3,0.6)[r]{$\pi$}}
\put(3.6,7.0){\circle*{0.3}}
\put(3.6,7.0){\line(2,1){2.8}}
\put(5.0,7.7){\vector(2,1){0.2}}
\put(6.8,8.15){\makebox(0.3,0.6)[l]{$q$}}
\put(3.6,7.0){\line(2,-1){2.8}}
\put(5.0,6.3){\vector(-2,1){0.2}}
\put(6.8,5.35){\makebox(0.3,0.6)[l]{$\bar q$}}
\put(3.0,4.8){\makebox(0.3,0.6)[c]{(a)}}
\put(1.0,3.4){\line(2,-1){2.8}}
\put(2.4,2.7){\vector(2,-1){0.2}}
\put(0.2,3.15){\makebox(0.3,0.6)[r]{$q$}}
\put(1.0,0.6){\line(2,1){2.8}}
\put(2.4,1.3){\vector(-2,-1){0.2}}
\put(0.2,0.35){\makebox(0.3,0.6)[r]{$\bar q$}}
\put(3.8,2.0){\circle*{0.3}}
\put(4.0,2.0){\pionh}
\put(6.8,1.75){\makebox(0.3,0.6)[l]{$\pi$}}
\put(4.0,0.0){\makebox(0.3,0.6)[c]{(b)}}
\put(10.0,6.0){\line(2,-1){2.8}}
\put(11.4,5.3){\vector(2,-1){0.2}}
\put(9.3,5.75){\makebox(0.3,0.6)[r]{$q$}}
\put(10.0,3.0){\line(2,1){2.8}}
\put(11.4,3.7){\vector(-2,-1){0.2}}
\put(9.3,2.75){\makebox(0.3,0.6)[r]{$\bar q$}}
\put(12.97,4.52){\oval(0.4,0.6)}
\put(12.99,4.50){\oval(0.4,0.6)}
\put(13.01,4.48){\oval(0.4,0.6)}
\put(13.3,4.6){\pionu}
\put(16.6,6.0){\makebox(0.3,0.6)[l]{$\pi$}}
\put(12.9,4.4){\glond}
\put(16.6,2.5){\makebox(0.3,0.6)[l]{$G$}}
\put(12.75,2.0){\makebox(0.3,0.6)[c]{(c)}}
\put(20.0,7.0){\glonh}
\put(19.7,6.75){\makebox(0.3,0.6)[r]{$G$}}
\put(23.83,7.0){\circle*{0.3}}
\put(23.5,7.0){\glonu}
\put(27.2,8.2){\makebox(0.3,0.6)[l]{$G$}}
\put(23.5,7.0){\glond}
\put(27.2,5.2){\makebox(0.3,0.6)[l]{$G$}}
\put(23.0,5.0){\makebox(0.3,0.6)[c]{(d)}}
\put(20.0,2.0){\glonh}
\put(19.7,1.75){\makebox(0.3,0.6)[r]{$G$}}
\put(23.8,2.0){\circle*{0.3}}
\put(23.7,2.0){\line(2,1){3.0}}
\put(25.1,2.7){\vector(2,1){0.2}}
\put(27.2,3.3){\makebox(0.3,0.6)[l]{$q$}}
\put(23.7,2.0){\line(2,-1){3.0}}
\put(25.1,1.3){\vector(-2,1){0.2}}
\put(27.2,0.2){\makebox(0.3,0.6)[l]{$\bar q$}}
\put(23.0,0.0){\makebox(0.3,0.6)[c]{(e)}}

\end{picture}

\caption[ ]{
Fragmentation and hadronization of pion.
 } \label{figpion}

\end{center}
\end{figure}
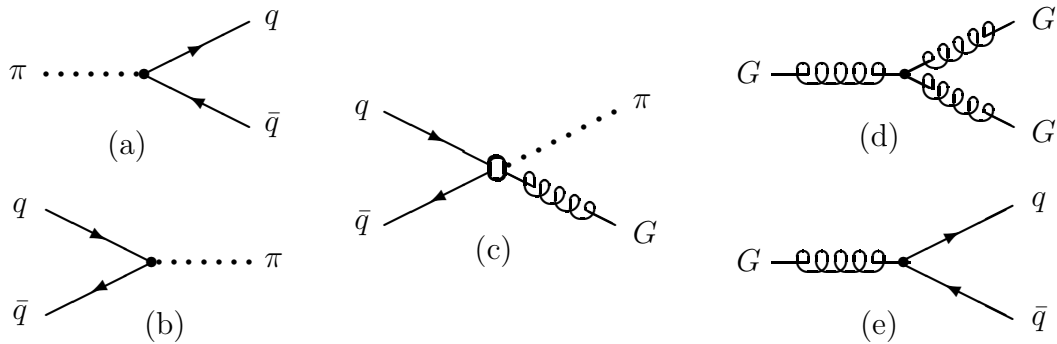
At high energy, pions may change to quark-gluon plasma phase.
If we consider only quarks, then it would corresponds to a
fragmentation of pion ($\pi$) into a quark($q$) and an antiquark
($\bar q$) which is depicted by Fig.\ref{figpion}(a).
A quark-antiquark system, in turn, clusterizes
into a pion as shown by Fig.\ref{figpion}(b).
This simplified case is quite similar to the nuclear multifragmentation
phenomena discussed in previous sections.
The number of particles are conserved for this case.

To consider new particle creation, we need to consider gluons ($G$) too
\cite{pion}, which may be created by the process of Fig.\ref{figpion}(c).
Due to the self-interaction of gluon, the gluon may break up itself
further giving more gluons, Fig.\ref{figpion}(d),
which is the same form as the branching phenomena of Fig.\ref{figbrnch}.
After the system has cooled down finally, there are only pions
without free quarks and gluons.
Thus we also need the annihilation process of gluon
($q\bar q$ pair production), Fig.\ref{figpion}(e).
The processes in Fig.\ref{figpion} may be viewed as a fragmentation
process of Fig.\ref{figbox} or as a branching process of Fig.\ref{figbrnch}.
The coefficient $C_k^l$ defined by Eq.({\ref{ckl}) becomes
$C^\pi_q$ and $C^\pi_{\bar q}$ with $C^\pi_q = C^\pi_{\bar q}$
for the process of $\pi \to q \bar q$, Fig.\ref{figpion}(a).
We also have $C_q^G = C_{\bar q}^G$  and $C_G^q = C_G^{\bar q}$.
Of course $C^\pi_q \ne C^q_\pi$ here; the rate of process (a) is
different from the rate of process (b) in general.
For the process of Fig.\ref{figpion}(d), the branching parameter
$\omega(t)$ of Eq.(\ref{furry}) may be assigned.

For the case of Fig.\ref{figpion},
we have four different types of particles (species $k$);
$\pi$, $q$, $\bar q$, and $G$.
Initially we have $n_\pi(0)$ pions without any quarks or gluons,
i.e., $n_q(0) = 0$, $n_{\bar q}(0) = 0$, and $n_G(0) = 0$.
The hot pions break up into $q \bar q$ pairs according to the process
(a), and then processes (c), (d), and (e) create more quarks and gluons.
In the intermediate step, the system may have all four species,
i.e., $n_\pi(t)$, $n_q(t)$, $n_{\bar q}(t)$, and $n_G(t)$ all
are nonzero in general with weighting parameters $x_\pi(t)$, $x_q(t)$,
$x_{\bar q}(t)$, and $x_G(t)$ respectively.
After the system has cooled down, we again have
$n_q(\infty) = n_{\bar q}(\infty) = n_G(\infty) = 0$
with $n_\pi(\infty) \ge n_\pi(0) \ne 0$.
If we put $x_q(\infty) = x_{\bar q}(\infty) = x_G(\infty) = 0$,
then we are secured to have
$n_q(\infty) = n_{\bar q}(\infty) = n_G(\infty) = 0$.
A more detailed study of this case would reveal a method of applying
our simple fragmentation model to a quark-hadron phase transition
including strangeness.

\section{Summary}   \label{concl}

\hspace*{30pt}
We have a simple exactly solvable model for nonequilibrium
fragmentation phenomena based on combinatorial analysis.
This model can describe nuclear multifragmentation and branching
phenomena.
The ubiquitous pairing phenomena is also able to be considered
with this model.

We have also looked at the similarities and differences of
quark-hadron phase transition to our fragmentation model.
But for a practical application of the model
to quark-hadron system, we need further studies,
such as what $\alpha_k$ and $\beta_k$ we should consider,
what $x_k(t)$ should be, how many particles ($N$) we need
to consider, etc.
If we consider different colors and flavors, the number of species
increases much more.
Including the strange quark, we might be able to
investigate the role of strangeness in a nuclear system.

It is a pleasure to acknowledge many discussions with Prof. Mekjian
during the course of various works on which this review is based.
I would also like to thank Prof. Bhang
for giving me the opportunity to participate this workshop.

\end{document}